\documentclass[aps,pra,twocolumn,floatfix,showpacs]{revtex4}
\usepackage{amsmath}
\usepackage{amsfonts}
\usepackage{amssymb}
\usepackage{graphicx}
\usepackage{color}

\begin{document}

\title{Optical memories and transduction of fields in double cavity optomechanical systems}
\author{Kenan Qu and G. S. Agarwal}
\affiliation{Department of Physics, Oklahoma State University, Stillwater, OK - 74078, USA}
\date{\today}

\begin{abstract}
Using electromagnetically induced transparency we show how the opto-mechanical systems can be used both as memory elements as well as for the transduction of optical fields. We use a double cavity optomechanical system which facilitates such applications. By tuning the frequency of the second cavity one can produce output fields at a variety of frequencies. We present analytical results for the steady state behavior which is controlled by the power and the detuning of the field driving the second cavity.
\end{abstract}
\maketitle

\section{Introduction}
The design of a good optical memory~\cite{network1,Gisin1,Gisin2} depends very much on the underlying physical process as well as system used to construct the memory. One needs systems or the storage elements with very long coherence times. The electromagnetically induced transparency (EIT)~\cite{EIT01,EIT02} has become an important physical mechanism to construct optical memories~\cite{storage1,storage2,storage3,storage4,Dey}. The optical pulses are stored in atomic coherences among long lived states. Many experiments have demonstrated the working of optical memories using typically atomic vapors. More recently the interest has shifted to other systems. The optomechanical systems (OMS) have very long coherence times and hence one has the possibility of using such systems as optical memories as information is stored in coherent phonons. It was proposed~\cite{Sumei1} and demonstrated experimentally~\cite{OMIT} that one has exact analog of EIT in OMS. Further phonons are generated coherently --- this being the analog of  the atomic coherence in  vapors. Thus EIT in OMS along with the long coherence time for the generated phonons can be used for making optical memories. A recent paper demonstrates the possibility of using OMS as a single photon router~\cite{router2,Sumei2}. There is also a closely related issue of the transduction of photons~\cite{router1} i.e. transferring photons of one frequency into another frequency~\cite{Tian,Wang}. This process is related to optical memories though there are demonstrations of such a process using the nonlinear process of frequency generation~\cite{Srinivasan}.

In this paper we focus on the optical memories and transduction of electromagnetic fields using OMS.
We propose the use of a double cavity OMS which provides one with considerable flexibility as far as the storage and retrieval of photons is concerned. The double cavity OMS also enables us to achieve transduction process to number of different frequencies including, in principle, the possibility of transduction from optical to microwave frequencies.

The organization of the paper is as follows. In section II we introduce the double cavity OMS~\cite{membrane1,membrane2} and present complete dynamical equations for this system. We obtain analytical solutions for the generated coherences and the field amplitudes in the steady state. The double cavity OMS can be used to control the EIT in a single cavity system. In section III we discuss the case when the coupling field (writing fields) and the probe field (signal field) are pulses. We show how the probe field is stored in coherent phonons and how the probe field can be retrieved by applying the reading pulse. We also show that the writing and reading processes are better with super Gaussian pulses~\cite{Dey}. In section IV we discuss distinct advantages of using the double cavity OMS. We discuss the transduction of fields.

\section{Basic Model and Equations for a double cavity optomechanical system}
\begin{figure}[phtb]
 \includegraphics[width=0.35\textwidth]{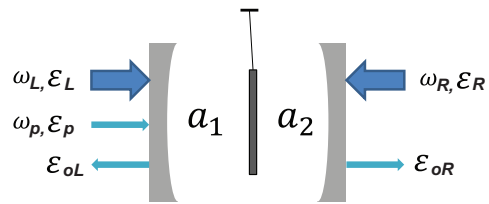}
 \caption{\label{Fig1}{(color online) Schematic double-cavity OMS.}}
\end{figure}

We study the system in Fig.~\ref{Fig1}, in which a mechanical resonator, coated with perfect reflecting films on both sides, is coupled to two cavities.
The mechanical resonator is modeled as a harmonic oscillator with mass $m$, frequency $\omega_m$ and momentum decay rate $\gamma_m$. For each of the cavity, we denote its field by $a_i$, frequency $\omega_i$ and decay rate $\kappa_i$, $i=1,2$. The field annihilation and creation operators satisfy the commutation relation $[a_i,a^\dag_j]=\delta_{ij}$. The probe pulse, or the probe with frequency $\omega_p$, is sent into cavity $1$. Our model is different from ``membrane in the middle'' setup~\cite{membrane1,membrane2}, because we are using an oscillating mirror which is $100\%$ reflective on both sides. The two cavities are coupled only via the oscillations of the mechanical mirror which are produced by the applied laser fields $\mathcal{E}_L$ and $\mathcal{E}_R$. Further the two cavities can be in different frequency regimes. We introduce normalized coordinates for the mechanical oscillator with commutation relations $[Q,P]=[x,p]/\hbar=\mathrm{i}$.
The Hamiltonian for the system as shown in Fig.~\ref{Fig1} is
\begin{align}\label{1}
    H=& \frac12\hbar\omega_m (P^2+Q^2)  + \sum_{i=1,2} (\hbar\omega_i a_i^\dag a_i -\hbar g_i a_i^\dag a_i Q ) \nonumber \\
      & + \mathrm{i}\hbar\mathcal{E}_L(a_1^\dag\mathrm{e}^{-\mathrm{i}\omega_Lt} - a_1\mathrm{e}^{\mathrm{i}\omega_Lt}) \nonumber \\
      & + \mathrm{i}\hbar\mathcal{E}_R(a_2^\dag\mathrm{e}^{-\mathrm{i}\omega_Rt} - a_2\mathrm{e}^{\mathrm{i}\omega_Rt}) \nonumber \\
      & + \mathrm{i}\hbar (a_1^\dag\mathcal{E}_p\mathrm{e}^{-\mathrm{i}\omega_p t} - a_1\mathcal{E}_p^*\mathrm{e}^{-\mathrm{i}\omega_p t}).
\end{align}
This Hamiltonian is to be supplemented by the dissipative terms corresponding to the Brownian motion of the mechanical oscillator and the leakage of the photons from the cavity. We will denote by $2\kappa_i$ this leakage rate for the $i$th cavity. Next we define $g_i,\mathcal{E}_L,\mathcal{E}_R,\mathcal{E}_p$ etc. Here $g_i$ is defined by $g_i=(\omega_{i}/L_i)\sqrt{\hbar/(2m\omega_m)}$ with $L_i$ being the length of the $i$th cavity. The $\mathcal{E}$ terms in (\ref{1}) denote the coupling of the cavity field to the applied laser fields. The coupling and probe fields are related to the power of the applied laser fields via  $\mathcal{E}_L = \sqrt{2\kappa_1P_L/(\hbar\omega_L)}$, $\mathcal{E}_R = \sqrt{2\kappa_2P_R/(\hbar\omega_R)}$ and $\mathcal{E}_p = \sqrt{2\kappa_1P_{p}/(\hbar\omega_p)}$, respectively. The term $H_\text{diss}$ denotes the damping of the two cavities and the mechanical resonator. We would work in the resolved-sideband regime, in which case $\omega_m\gg\kappa_i$.

It is convenient to remove all the fast frequencies $\omega_L,\omega_R,\omega_p$ from the Hamiltonian by redefining $a$'s as $a_1\to a_1\mathrm{e}^{-\mathrm{i}\omega_Lt}$, $a_2\to a_2\mathrm{e}^{-\mathrm{i}\omega_Rt}$, then the Hamiltonian becomes
\begin{align}\label{2}
    H =& \frac12\hbar\omega_m (P^2+Q^2) - \sum_{i=1,2} \hbar g_i a_i^\dag a_i Q \nonumber \\
      & + \hbar(\omega_1-\omega_L) a_1^\dag a_1 + \hbar(\omega_2-\omega_R) a_2^\dag a_2 \nonumber \\
      & + \mathrm{i}\hbar\mathcal{E}_L(a_1^\dag - a_1) + \mathrm{i}\hbar\mathcal{E}_R(a_2^\dag - a_2) \nonumber \\
      & + \mathrm{i}\hbar (a_1^\dag\mathcal{E}_p\mathrm{e}^{-\mathrm{i}\delta t} - a_1\mathcal{E}_p^*\mathrm{e}^{\mathrm{i}\delta t}). \nonumber \\
    \delta =& \omega_p-\omega_L
\end{align}

Using Eq. (\ref{2}) we can derive the quantum Langevin equations for the operators $Q$, $P$, $a_i$ and $a_i^\dag$. However, for the purpose of the present paper, we would work with semiclassical equations so that all operator expectation values are replaced by their mean values. Thus in the rest of the paper, all the quantities $Q$, $P$, $a_i$ and $a_i^\dag$ would be numbers. The equations of motion for these are
\begin{align}
		\dot{Q} & =\omega_{m}P, \label{3}\\
		\dot{P} & =(g_{1}a_{1}^{\dagger}a_{1}+g_{2}a_{2}^{\dagger}a_{2})-\omega_{m}Q-\gamma_{m}P, \label{4}\\
		\dot{a}_{1} & = -\mathrm{i}(\omega_1-\omega_L-g_1Q)a_{1} -\kappa_{1}a_{1} +\mathcal{E}_L +\mathcal{E}_{p}\mathrm{e^{-\mathrm{i\delta t}}}, 	\label{5}	\\
		\dot{a}_{2} & = -\mathrm{i}(\omega_2-\omega_R+g_2Q)a_{2}-\kappa_{2}a_{2}+\mathcal{E}_R.  \label{6}
\end{align}
Note that the Eqs. (\ref{3})-(\ref{6}) involve periodically oscillating terms and hence in the long time limit, any of the fields and the mechanical coordinates will have a solution of the form $A=\sum_{n=-\infty}^{+\infty} \mathrm{e}^{-\mathrm{i}n\delta t}A_n$. The $A_n$'s can be obtained by the Floquet analysis.

We assume that the probe is much weaker than the coupling field, then $A_n$'s can be obtained perturbatively. In particular, we find the steady state results to first order in $|\mathcal{E}_{p}/\mathcal{E}_{L,R}|$
\begin{align}
		a_{10} &= \frac{\mathcal{E}_L}{\kappa_1+\mathrm{i}\Delta_1}, \qquad a_{20} = \frac{\mathcal{E}_R}{\kappa_2+\mathrm{i}\Delta_2}, \label{7}\\
        Q_0 &= \frac{1}{\omega_m}(g_1|a_{10}|^2-g_2|a_{20}|^2), \label{8}\\
		Q_+ &= -\frac{1}{d(\delta)}\frac{g_1a_{10}^*\mathcal{E}_{p}}{(\kappa_1+\mathrm{i}\Delta_1 -\mathrm{i}\delta)}, \label{9}\\
		d(\delta) &= \sum_{i=1,2} \frac{2\Delta_ig_i^2|a_{i0}|^2}{(\kappa_i-\mathrm{i}\delta)^2 + \Delta_i^2} - \frac{\omega_m^2-\delta^2-\mathrm{i}\delta\gamma_m}{\omega_m}, \label{10} \\
		a_{1+} &= \frac{\mathrm{i}g_1a_{10}}{(\kappa_1+\mathrm{i}\Delta_1-\mathrm{i}\delta)}Q_+ + \frac{\mathcal{E}_p}{(\kappa_1+\mathrm{i}\Delta_1-\mathrm{i}\delta)}, \label{11} \\
		a_{1-} &= \frac{\mathrm{i}g_1a_{10}}{(\kappa_1+\mathrm{i}\Delta_1+\mathrm{i}\delta)}Q_+^*, \label{12} \\
    	a_{2+} &=\frac{-\mathrm{i}g_2a_{20}}{(\kappa_2+\mathrm{i}\Delta_2-\mathrm{i}\delta)}Q_+,		 \label{13}\\
    	a_{2-} &=\frac{-\mathrm{i}g_2a_{20}}{(\kappa_2+\mathrm{i}\Delta_2+\mathrm{i}\delta)}Q_+^*,		 \label{14}
\end{align}
where $\Delta_1=\omega_1-\omega_L-g_1Q_0$ and $\Delta_2=\omega_2-\omega_R+g_2Q_0$ are the detuning of the coupling laser to the effective cavity frequencies. The $Q_0$ denotes the displacement of the mechanical resonator under radiation pressure. The cavity field $a_{i0}$ is the field in the $i$th cavity at the frequency of the coupling laser. The field $a_{i\pm}$ is the field at the frequency $\omega_{L,R}+\delta = \omega_{L,R}+(\omega_p \pm \omega_{L,R})$, and specifically $\omega_{1+}=\omega_p$. The output fields are defined as
\begin{align}
    \mathcal{E}_{oL} &= 2\kappa_1(a_{1+}\mathrm{e}^{-\mathrm{i}(\omega_L+\delta) t} + a_{1-}\mathrm{e}^{-\mathrm{i}(\omega_L-\delta) t}) - \mathcal{E}_p\mathrm{e}^{-\mathrm{i}\omega_p t} \label{15}\\
    \mathcal{E}_{oR} &= 2\kappa_2(a_{2+}\mathrm{e}^{-\mathrm{i}(\omega_R+\delta) t} + a_{2-}\mathrm{e}^{-\mathrm{i}(\omega_R-\delta) t}) \label{16}
\end{align}
Note that the component $a_{2+}$ would yield the output at the frequency $\omega_R+\omega_p-\omega_L$ whereas the component $a_{2-}$ produces an output at the frequency $\omega_R-\omega_p+\omega_L$.
We would typically consider the situation when $\omega_p$ is close to the cavity frequency. The field $\omega_L$ and $\omega_p$ combine to produce phonons at the frequency $\omega_p-\omega_L \approx \omega_m$, i.e. $Q_+\neq0$. This is the reason for the production of coherent phonons. For $\mathcal{E}_{R}=0$, the Eqs. (\ref{7})-(\ref{14}) lead to the well-known results for the single cavity. In particular the equation (\ref{11}) in the limit of $\mathcal{E}_{R}=0$ leads to the EIT in OMS~\cite{EIT01,EIT02}. The EIT is the reason that OMS can be used as optical memory elements~\cite{Sumei1}.

\section{Storage and Retrieval of Optical Pulse in Single Cavity OMS}
We now discuss how OMS can be used as optical memory elements. The first demonstration of this was given by Fiore et al~\cite{Wang}. We first note that Eq. (\ref{11}) ($\mathcal{E}_{R}=0$) leads to a width of the EIT window given by
\begin{equation}\label{17}
    \Gamma_\text{EIT} \approx \frac{\gamma_m}{2}+\frac{g_1^2}{\kappa_1}|a_{10}|^2.
\end{equation}
\begin{figure}[tbp]
 \includegraphics[width=0.45\textwidth]{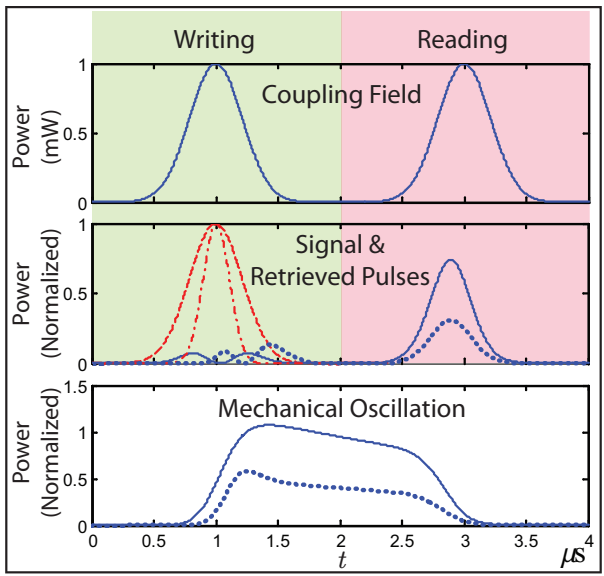}
 \caption{\label{Fig2}{(color online) Numerical simulation of ``writing'' and ``reading'' Gaussian probe pulses in a single cavity OMS. We plot two probe pulses with different width $\tau_p=0.15\mu$s (red dot-dashed curve) and $\tau_p=0.3\mu$s (red dashed curve). The Gaussian coupling pulses have width $\tau_L=0.3\mu$s and their peak power is $P_L=1$mW. The powers of the output pulses $|(2\kappa_1a_{1+}(t)-\mathcal{E}_{p}(t))/\mathcal{E}_{p}|^2$ and the mechanical oscillation $|\kappa_1Q_+(t)/\mathcal{E}_{p}|^2$ are normalized to the peak power of the probe pulse, which is much less than the coupling pulses. In the middle and bottom panels, the blue solid curves illustrate the result corresponding to probe pulse with $\tau_p=0.15\mu$s, and the blue dotted curves corresponding to $\tau_p=0.3\mu$s.}}
\end{figure}
Clearly, if the input probe field is a pulse, then its spectral width has to be less than $\Gamma_\text{EIT}$ in order to have distortionless propagation of the probe pulse. We next examine propagation of pulses using Eqs. (\ref{3})-(\ref{6}). We substitute solutions in the form $\sum_n \mathrm{e}^{-\mathrm{i}n\delta t}A_n(t)$ and obtain time dependent equations for $A_n(t)$. These equations are truncated by assuming that $|\mathcal{E}_p(t)| \ll |\mathcal{E}_{L}(t)|$. For $\mathcal{E}_p(t)$ and $\mathcal{E}_{L}(t)$, we take Gaussian shapes
\begin{align}
     \mathcal{E}_p(t) &= \mathcal{E}_p \mathrm{e}^{-\frac{(t-t_{wr})^2}{2\tau_p^2}}, \label{22}\\
     \mathcal{E}_{L}(t) &= \mathcal{E}_{L} \mathrm{e}^{-\frac{(t-t_{wr})^2}{2\tau_L^2}} + \mathcal{E}_{L} \mathrm{e}^{-\frac{(t-t_{rd})^2}{2\tau_L^2}}, \label{23}
\end{align}
where $t_{wr}$ and $t_{rd}$ are the central time of the ``writing'' and ``reading'' coupling lasers. We also assume that the width $\tau_L$ of the coupling laser is no less than the width $\tau_p$ of the probe pulse. Further, we assume that $\tau_p^{-1} < \Gamma_\text{EIT}$. We use the Fourth-order Runge-Kutta method to solve the time dependent equations with Eqs. (\ref{22}) and (\ref{23}) as the initial conditions.

We first examine the case when $P_{R}=0$. For numerical simulations, we use the following parameters $m=20$ng, $g/2\pi=1.55$kHz, $\gamma_m/2\pi=41$kHz, $\omega_m/2\pi=51.8$MHz, $\kappa/2\pi=1.5$MHz, $\lambda=775$nm. It should be borne in mind that Eqs.~(\ref{3})-(\ref{6}) are nonlinear and hence we have to ensure that the parameters are such that the solutions to (\ref{3})-(\ref{6}) stable. We have checked this using the Routh-Hurwitz criterion~\cite{Hurwitz}. We also assume that the coupling laser is red detuned with respect to the cavity frequency $\omega_1$, i.e. $\Delta_1=\omega_1-\omega_l-g_1Q_0\approx \omega_m$ and that the probe has a frequency in resonance with the cavity $\omega_p=\omega_1$.
\begin{figure}[tbp]
 \includegraphics[width=0.45\textwidth]{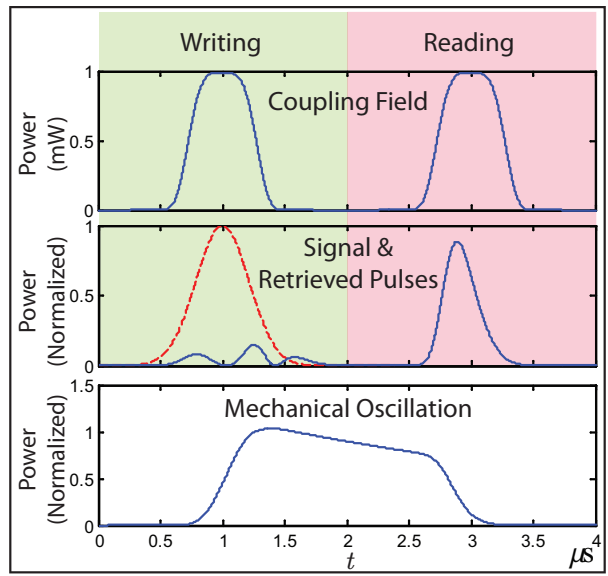}
 \caption{\label{Fig3}{(color online) Numerical simulation of the protocol of optical memory using single-cavity OMS. Curves are defined similar to Fig.~\ref{Fig2}, except the coupling pulses are super-Gaussian and $\tau_L=\tau_p=0.3\mu$s.}}
\end{figure}
\begin{figure*}[tpb]
 \includegraphics[width=0.95\textwidth]{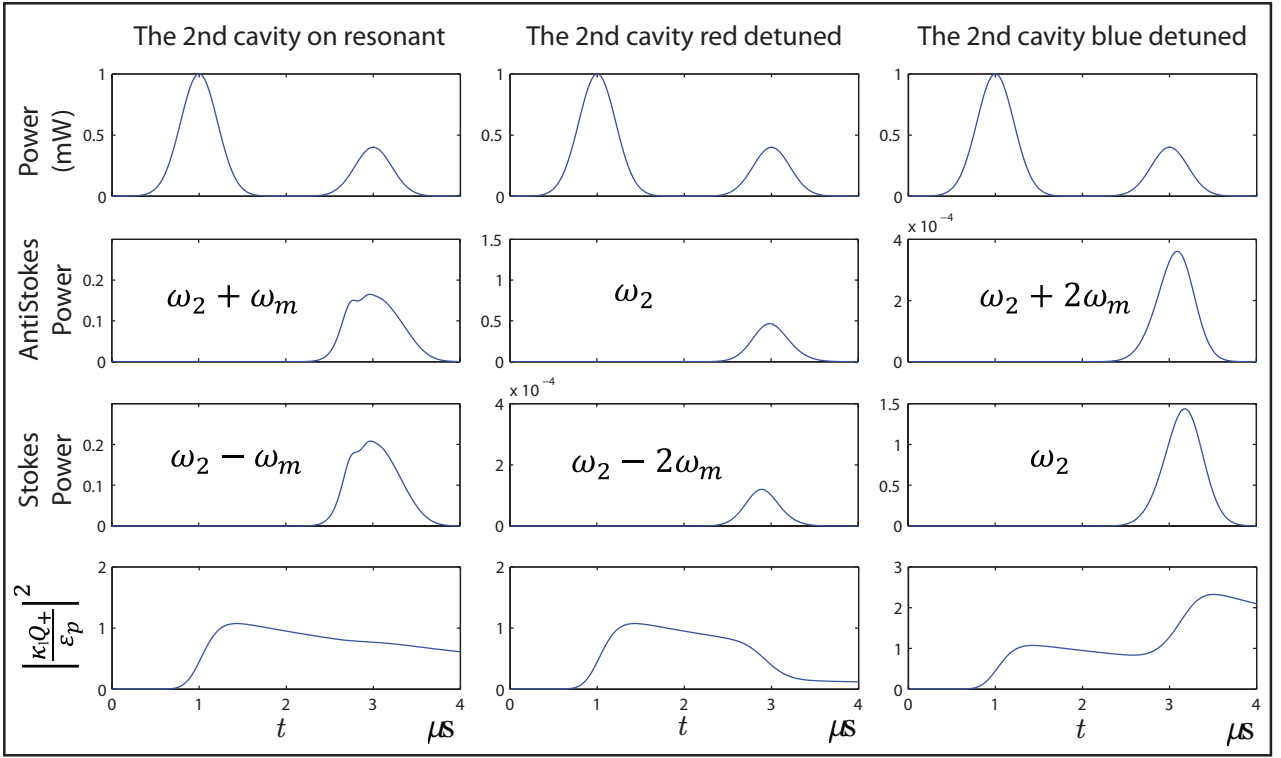}
 \caption{\label{Fig4}{(color online) Numerical simulation of the optical memory using single-cavity OMS when the ``reading'' cavity frequency $\omega_2$ is on resonant, red detuned and blue detuned with the reading laser frequency $\omega_R$. The peak power is $P_L=1$mW for the writing coupling laser and $P_R=0.4$mW for the reading coupling laser.}}
\end{figure*}
\begin{figure*}[btp]
 \includegraphics[width=0.95\textwidth]{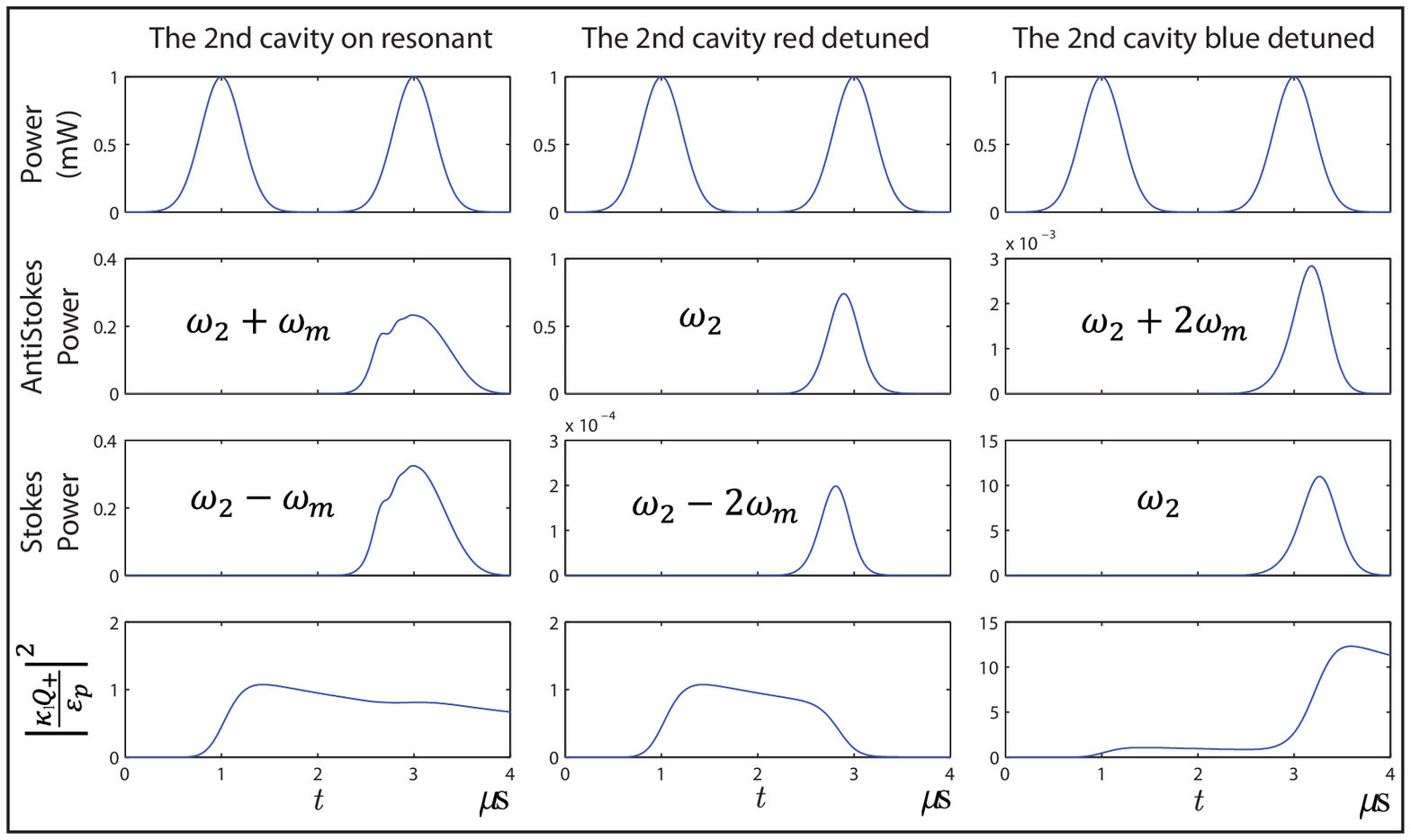}
 \caption{\label{Fig5}{(color online) Numerical simulation of the optical memory using single-cavity OMS. Curves are defined similar to Fig.~\ref{Fig4} except the peak power is $P_R=1$mW for the reading laser.}}
\end{figure*}

We show two typical sets of numerical simulation results in Fig.~\ref{Fig2}. In the ``coupling'' stage, we send in both the coupling laser and the probe probe pulse simultaneously. The width of the coupling laser is $\tau_L=0.3\mu$s; and the widths of the probe pulses are $\tau_p=0.3\mu$s (dashed curve and solid curve as corresponding result) and $\tau_p=0.15\mu$s (dot-dashed curve and dashed curve as corresponding result). Assuming they are both Fourier-limited Gaussian pulses which have time-bandwidth product $\sim 0.44$, their spectral widths can be calculated as $\Delta\omega=0.44/\tau_p=2\pi0.47$MHz and $2\pi0.23$MHz for $\tau_p=0.15\mu$s and $0.3\mu$s, respectively.  The peak power of the coupling pulse is $P_L=1$mW. It produces an EIT window with width $\Gamma_\text{EIT}=2\pi11$MHz (Eq.(\ref{17})) which is much wider than the spectrum widths of the probe pulses. The optical field in the probe pulses are converted into coherent phonons of the mechanical oscillator as shown in the bottom panel of Fig.~\ref{Fig1}. This is because of the coherent process $\omega_p-\omega_L= \omega_m$. The coherent phonon survives over a time scale of the order of $\gamma_m^{-1}$ which is much longer than the cavity lifetime $\kappa_1^{-1}$. The probe pulse can be retrieved by applying the ``reading'' pulse at a time later (within $\gamma_m^{-1}$). The application of the ``reading'' pulse converts the coherent phonons into light field via the upconversion process $\omega_L+\omega_m \to \omega_p=\omega_1$. Fiore et al.~\cite{Wang} demonstrated the storage and retrieval of light pulses.

We observe that the probe pulse with a larger temporal width $\tau_p=0.3\mu$s ($\Gamma_\text{EIT}\tau_p=21$) is stored better than the pulse with width $\tau_p=0.15\mu$s ($\Gamma_\text{EIT}\tau_p=10$). The conversion to phonons takes more efficiently for $\tau_p=0.3\mu$s. This then results in better retrieval of the probe pulse. The retrieved peak powers are $0.74|\mathcal{E}_p|^2$, $0.31|\mathcal{E}_p|^2$ for $\tau_p=0.3\mu$s and $\tau_p=0.15\mu$s, respectively.

Earlier work with atomic systems~\cite{Dey} has shown that the storage and retrieval processes are more efficient if the Gaussian pulses are replaced by super-Gaussian pulses.
\begin{equation}\label{18}
     \mathcal{E}_{L}(t) = \mathcal{E}_{L} \mathrm{e}^{-\frac12\left(\frac{t-t_{wr}}{\tau_L}\right)^\beta} + \mathcal{E}_{L} \mathrm{e}^{-\frac12\left(\frac{t-t_{rd}}{\tau_L}\right)^\beta}.
\end{equation}
For $\beta=4$, we have adiabatic switching on and off of the coupling fields. It has a more rectangular tempo profile with sharp edges. Fig.~\ref{Fig3} shows the result of numerical simulation using super-Gaussian shaped ``writing'' and ``reading'' laser pulses. Comparing Figs.~\ref{Fig2} and \ref{Fig3}, we find the super-Gaussian coupling pulses produce a retrieved pulse with sharper front edge and higher peak power $0.79|\mathcal{E}_p|^2$.

The almost complete recovery of the weak probe pulse is especially significant in the context of single photon optical memories.

\section{Storage, Retrieval and transduction of fields in a Double Cavity Optomechanical System}
The double-cavity OMS bring a larger flexibility to the optical pulse storage protocol and applications. One can optimize these two independent cavities for their own functions with different parameters, like cavity decay rate, resonance frequency and coupling rate. We display in Figs.~\ref{Fig4} and~\ref{Fig5} a series of output fields when the second cavity is red detuned, on resonance and blue detuned. We take $\kappa_2=\kappa_1$ though additional flexibility in the operation of the memory device is possible by making them different. The fields at the output of the second cavity $\mathcal{E}_{oR}$ have Stokes and antiStokes components, whose central frequencies are given by $\omega_R\pm\omega_m$. They are related the amplitudes $a_{2+}$ and $a_{2-}$, respectively, by Eq.~(\ref{16}).

When the second cavity is red detuned, the antiStokes pulse is on resonance with the second cavity whereas the Stokes pulse is far off resonance. This is the reason of very little Stokes output. The antiStokes output can be comparable to the input probe pulse $\mathcal{E}_p$ depending on the power used to pump the second cavity. With higher applied power, the conversion of phonons to the antiStokes field is more efficient. If the second cavity is on resonance, then, as expected, the generated Stokes and antiStokes are of comparable magnitude~\cite{footnote}. These curves clearly show that using the second cavity on resonance produces better coherent outputs than the case when the second cavity is red detuned. Note that the incident probe has a frequency $\omega_p=\omega_L+\omega_m\approx\omega_1$ whereas the outputs from the second cavity have frequencies $\omega_2\pm\omega_m$. We have here phonon induced transduction of photon fields from a frequency $\omega_1$ to frequency $\omega_2\pm\omega_m$. We also produce two outputs.

For the case of the blue detuning of the second cavity $\omega_R\approx\omega_2+\omega_m$, the generated antiStokes field is far off resonance whereas the Stokes field is on resonance. Therefore very significant amount of the Stokes field is generated. Further the blue detuned laser leads to the generation of the coherent phonons as shown by the Fig.~\ref{Fig4}. The nonlinear mixing process involving the field at $\omega_R$ and phonons at $\omega_m$ produce the Stokes field at $\omega_2$. The increase of the phonon excitation can be understood from a quantum mechanical description of the process---the radiation matter interaction for the second cavity is
\begin{align}\label{31}
    & (a_R\mathrm{e}^{-\mathrm{i}\omega_Rt} + a_2\mathrm{e}^{-\mathrm{i}\omega_2t})^\dag (a_R\mathrm{e}^{-\mathrm{i}\omega_Rt} + a_2\mathrm{e}^{-\mathrm{i}\omega_2t}) \nonumber \\
    &\times (Q_+\mathrm{e}^{-\mathrm{i}\omega_mt} + Q_+^\dag\mathrm{e}^{\mathrm{i}\omega_mt}) \nonumber \\
    =& (Q_+^\dag a_2^\dag a_R + H.c.) + \text{non resonant terms}.
\end{align}
This clearly shows how a photon of frequency $\omega_R$ gets converted into a phonon and a photon of frequency $\omega_2$. In Fig.~\ref{Fig5}, we show the output fields from the second cavity when the field driving the second cavity is large. The idea here is to see how well a very weak pulse applied at the frequency $\omega_p$ from the left would be recovered. The Fig.~\ref{Fig5} shows that the recovery is good. This should be especially relevant for the transduction of single photons. These results show how the coherently generated phonons in the first cavity can be used for the transduction of the optical fields. In case if the second cavity were a microwave cavity driven by microwave field, then one has the possibility of converting incident optical fields into microwave fields.

\section{Conclusion}
In conclusion we have shown how single and double cavity opto-mechanical systems can be used for optical memories and for transduction of the optical fields. We developed the mathematical model and analyzed it both numerically and analytically. As in the case of EIT in atomic vapors, the use of super Gaussian writing pulses is advantageous. We presented results for the dynamics of phonons which are generated coherently. The dynamics of phonons shows clearly how the fields are stored and how these could be converted back into fields with frequencies which depend on the power and the detuning of the fields driving the second cavity. The extension of these results to quantized signal or probe pulses~\cite{clerk,Kippenberg} would be interesting and would be presented elsewhere.

\end{document}